\PassOptionsToPackage{unicode}{hyperref}
\PassOptionsToPackage{hyphens}{url}
\documentclass[
]{article}
\usepackage{amsmath,amssymb}
\usepackage{iftex}
\ifPDFTeX
  \usepackage[T1]{fontenc}
  \usepackage[utf8]{inputenc}
  \usepackage{textcomp} % provide euro and other symbols
\else % if luatex or xetex
  \usepackage{unicode-math} % this also loads fontspec
  \defaultfontfeatures{Scale=MatchLowercase}
  \defaultfontfeatures[\rmfamily]{Ligatures=TeX,Scale=1}
\fi
\usepackage{lmodern}
\ifPDFTeX\else
\fi
\IfFileExists{upquote.sty}{\usepackage{upquote}}{}
\IfFileExists{microtype.sty}{% use microtype if available
  \usepackage[]{microtype}
  \UseMicrotypeSet[protrusion]{basicmath} % disable protrusion for tt fonts
}{}
\makeatletter
\@ifundefined{KOMAClassName}{% if non-KOMA class
  \IfFileExists{parskip.sty}{%
    \usepackage{parskip}
  }{% else
    \setlength{\parindent}{0pt}
    \setlength{\parskip}{6pt plus 2pt minus 1pt}}
}{% if KOMA class
  \KOMAoptions{parskip=half}}
\makeatother
\usepackage{xcolor}
\usepackage{longtable,booktabs,array}
\usepackage{calc} % for calculating minipage widths
\usepackage{etoolbox}
\makeatletter
\patchcmd\longtable{\par}{\if@noskipsec\mbox{}\fi\par}{}{}
\makeatother
\IfFileExists{footnotehyper.sty}{\usepackage{footnotehyper}}{\usepackage{footnote}}
\makesavenoteenv{longtable}
\usepackage{graphicx}
\makeatletter
\def\maxwidth{\ifdim\Gin@nat@width>\linewidth\linewidth\else\Gin@nat@width\fi}
\def\maxheight{\ifdim\Gin@nat@height>\textheight\textheight\else\Gin@nat@height\fi}
\makeatother
\setkeys{Gin}{width=\maxwidth,height=\maxheight,keepaspectratio}
\makeatletter
\def\fps@figure{htbp}
\makeatother
\usepackage[margin=1in]{geometry}
\usepackage{authblk}
\usepackage{caption}
\usepackage{float}
\usepackage{xurl}
\ifLuaTeX
  \usepackage{selnolig}  % disable illegal ligatures
\fi
\IfFileExists{bookmark.sty}{\usepackage{bookmark}}{\usepackage{hyperref}}
\IfFileExists{xurl.sty}{\usepackage{xurl}}{} % add URL line breaks if available
\hypersetup{
  pdftitle={Direct Neutrino Communication Through the Earth: NuMI-Calibrated Simulation and Far-Field Sensitivity},
  pdfauthor={Dr. Sven-Patrik Hallsjö},
  hidelinks,
  pdfcreator={LaTeX via pandoc}}

\title{Direct Neutrino Communication Through the Earth: NuMI-Calibrated
Simulation and Far-Field Sensitivity}
\author[1]{Dr. Sven-Patrik Hallsjö\thanks{\texttt{patrik.hallsjo@gmail.com}}}
\affil[1]{Independent researcher, Stockholm, Sweden}
\date{Dated: September 2026}

\begin{document}
\maketitle

\hypertarget{abstract}{%
\section{Abstract}\label{abstract}}

Neutrinos can pass through substantial matter, potentially enabling
direct links where electromagnetic paths are obstructed, but their weak
interactions make reception difficult. We present two separate
calculations. First, a repeated OOK/CRC simulation calibrated to the
published NuMI--MINERvA selected-event mean correctly accepts 65.7\% of
simulated packets with a 40-bit payload and an eight-bit CRC after five
repetitions, with about 587 s mean latency under assumed external
synchronization. Varying the input mean across its approximate Poisson
counting interval gives 60--72\% correct acceptance; this is a
sensitivity range, not a joint confidence interval. Second, a distinct
idealized far-field model gives 24.4 MW peak neutrino-carried on-slot
power for a 10 kt receiver at 5,000 km, conditional on 3 GeV neutrinos,
1 mrad divergence, 50\% selection efficiency, unit flavor survival, zero
background and one-second slots. Neither result establishes a practical
long-baseline source or measured CRC performance.

\textbf{Keywords:} neutrino communication; through-Earth links; Poisson
channel; accelerator beam; detector sensitivity; feasibility.

\begin{figure}
\centering
\includegraphics{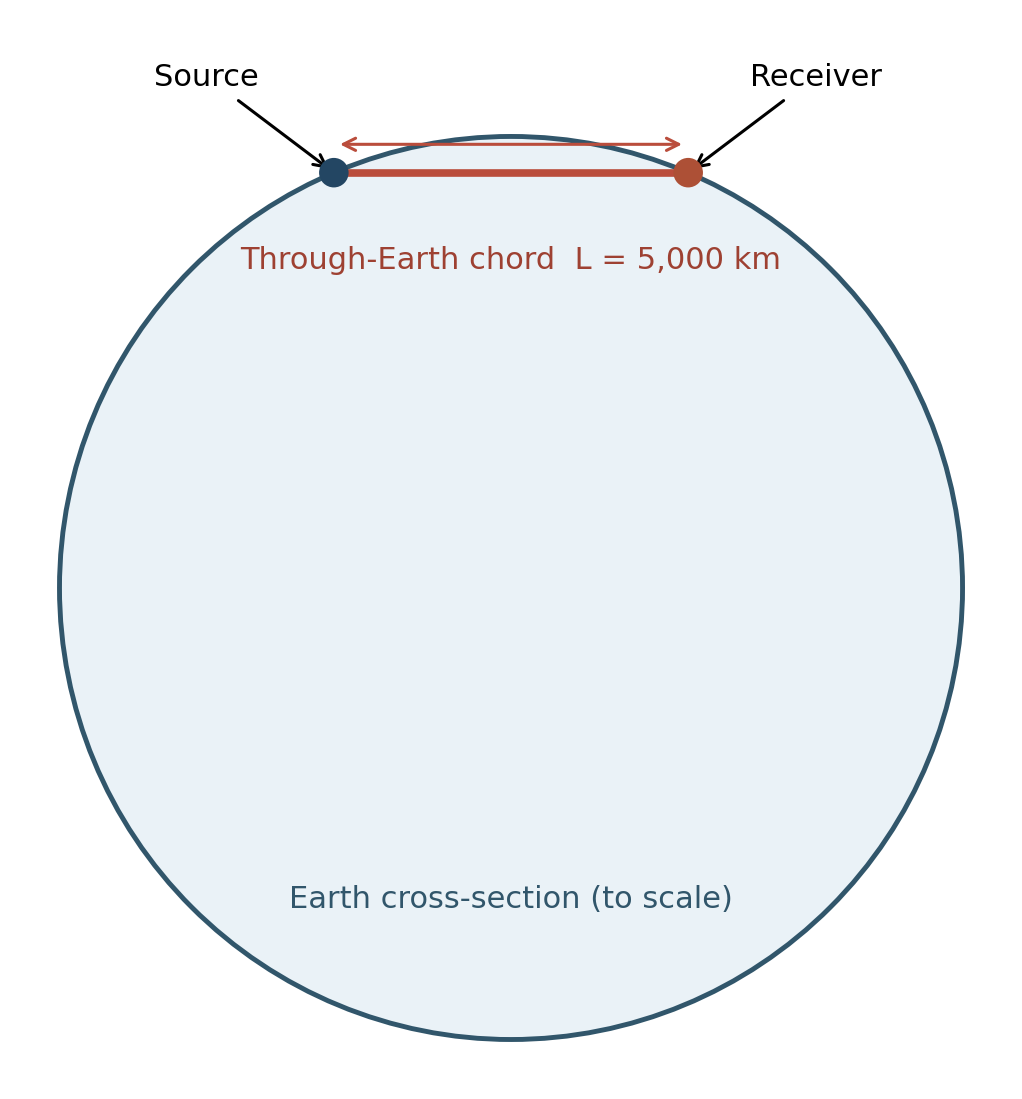}
\caption{To-scale Earth cross-section and a 5,000 km through-Earth
chord. The beam width is omitted; the diagram does not depict angular
divergence. A fixed-source example used to define chord length. The
source and receiver lie on the surface; a real installation would
require a beamline oriented along the chord.}
\end{figure}

\hypertarget{introduction}{%
\section{1. Introduction}\label{introduction}}

Communication through rock or across a large terrestrial chord
ordinarily requires an indirect route, such as a surface relay or
infrastructure that crosses the barrier. A neutrino beam offers a
physically distinct direct path because most neutrinos traverse matter
without interacting. This property could matter in a narrowly defined
setting where a direct path has operational value and an electromagnetic
relay is unavailable, obstructed or too slow to deploy. It does not make
neutrinos generally preferable: their weak interactions mean that only a
small fraction of transmitted particles produce identifiable events, so
a detectable beam need not deliver a useful message at acceptable rate,
energy cost or receiver scale.

The research motivation is therefore quantitative. Earlier work
established the concept and a short experimental link, but those results
do not by themselves show how measured event statistics translate into a
finite message protocol, or how the event requirement scales when
baseline, beam spread and detector mass change. This scoping study makes
those steps explicit under deliberately idealized assumptions, so that
the dominant penalties and missing physical inputs are visible before
anyone treats an application example as an engineering proposal. Its
purpose is to delimit the question and identify what a source-specific
study would need to calculate; it does not argue that a practical use
case has already been found.

The possibility has passed an experimental proof-of-principle test.
Stancil \emph{et al.} transmitted an encoded message using the NuMI beam
and MINERvA detector over 1.035 km, including 240 m of earth. They
report a decoded rate of 0.1 bit s⁻¹ and a 1\% bit error rate {[}1{]}.
That result establishes a link at its particular source, geometry,
detector and decoding protocol; it does not answer whether a regional or
global direct link could be operated at a useful rate.

Huber studied a one-way link to a submerged submarine using a
high-energy neutrino beam from a muon storage ring {[}2{]}. Learned,
Pakvasa and Zee examined signalling on galactic scales {[}3{]}. A later
submarine-navigation proposal addresses position, navigation and timing
rather than reliable delivery of an arbitrary message {[}4{]}. These
studies motivate a clear separation between demonstrated transmission
and projections under particular technologies.

This study also follows Hallsjö's analysis of locating nuclear-powered
submarines by their emitted antineutrinos {[}5{]}. That work treated an
uncontrolled, comparatively low-energy reactor source as a detection
problem and examined sparse counts, background, geometry and detector
deployment. Here the question is different: for a controlled and
modulated source, what signal count is needed to decode bits, and how
does an idealized source requirement change with baseline, angular
spread and receiver size? The detector and interaction expertise
documented in Hallsjö's Baby MIND thesis informs the next,
detector-specific stage of this work {[}6{]}. Reactor-antineutrino event
rates or detector assumptions from {[}5{]} are not transferred to the
high-energy communication example.

Global neutrino telecommunications were already examined quantitatively
by Sáenz \emph{et al.} in 1977 {[}16{]}. The present novelty therefore
cannot be the existence of that idea. Our aim is a transparent
\emph{scoping calculation} that connects an explicit error target to
detector size and neutrino-carried energy rate, adds a finite-message
example calibrated at the one measured source--receiver geometry, and
tests the scaling against proposed uses with distinct receiver and
latency constraints. We report what the stated assumptions imply and
identify which missing inputs prevent an engineering conclusion.

\hypertarget{link-definition-and-communication-metric}{%
\section{2. Link definition and communication
metric}\label{link-definition-and-communication-metric}}

Consider a stationary source that sends on--off-keyed symbols toward a
stationary receiver through a chord of length \(L\). In an ideal
synchronized slot, an on symbol yields a selected signal count with
Poisson mean \(s\); an off symbol has no signal. A background count with
mean \(b\) may be present in either slot. Conditional on transmitted
symbol \texttt{x\ ∈\ \{0,1\}}, the count model is

\[
K | x \sim \operatorname{Poisson}(b + s x) \tag{1}
\]

For equally probable input symbols, \(b = 0\), and a threshold of one
observed event, false-positive errors vanish and an on symbol is missed
with probability \(\exp(-s)\). Therefore

\[
BER_uncoded = \frac{1}{2} \exp(−s) \tag{2}
\]

This equation measures uncoded, synchronized \emph{raw} bits. A
delivered payload rate must additionally account for symbol slots,
synchronization, framing, coding, source duty cycle and any failed
frames. An actual background \(b > 0\) introduces both false-positive
and false-negative errors; neither Eq. (2) nor a threshold of one should
then be assumed optimal.

We choose a target \(BER_uncoded = 0.01\), so the mean signal
requirement under the stated ideal conditions is

\[
s_required = −\ln(2 × 0.01) = 3.912 \text{selected events per on symbol} \tag{3}
\]

\begin{longtable}[]{@{}
  >{\raggedright\arraybackslash}p{(\columnwidth - 2\tabcolsep) * \real{0.5000}}
  >{\raggedright\arraybackslash}p{(\columnwidth - 2\tabcolsep) * \real{0.5000}}@{}}
\toprule\noalign{}
\begin{minipage}[b]{\linewidth}\raggedright
Quantity
\end{minipage} & \begin{minipage}[b]{\linewidth}\raggedright
Definition used here
\end{minipage} \\
\midrule\noalign{}
\endhead
\bottomrule\noalign{}
\endlastfoot
Slot duration, \({\tau}\) & Time allocated to one OOK symbol, whether 0
or 1 \\
Slot rate, \(R_slot\) & \(1/{\tau}\) slots s⁻¹; for uncoded OOK, this
equals raw input bits s⁻¹ \\
Peak on-slot power, \(P{\nu},on\) & Neutrino-carried energy required in
a 1 slot, divided by \({\tau}\) \\
Full-duty average, \(P{\nu},full\) & \(P{\nu},on\) if every slot
transmits a 1 \\
Time-average power, \(P{\nu},avg\) & \(q_{1}P{\nu},on\), where \(q_{1}\)
is the fraction of slots carrying a 1; \(P{\nu},on/2\) for equiprobable
OOK \\
\end{longtable}

The far-field rate model sets \({\tau} = 1/R_slot\), so increasing the
bit rate at fixed receiver and divergence requires a proportionally
greater peak on-slot power. A coded scheme must instead define its coded
slot rate, code rate and symbol energy explicitly.

Let \({\tau}\) be the duration of one OOK slot, \(R_slot = 1/{\tau}\)
the slot rate, and \(q_{1}\) the fraction of slots carrying a one. With
one uncoded bit per slot, the raw bit rate is \(R_raw = R_slot\). Let
\(E{\nu},1 = s E{\nu}/p_sel\) be the neutrino-carried energy needed in
each on slot. The peak power while transmitting a one is
\(P{\nu},on = E{\nu},1/{\tau}\). If the beam transmits in every slot,
this is also its full-duty average power. For random, equiprobable
uncoded OOK, the time-averaged neutrino-carried power is
\(P{\nu},avg = q_{1}P{\nu},on = P{\nu},on/2\). The numerical tables
state \textbf{peak on-slot power} unless they explicitly say otherwise.
FEC, framing and nonuniform source scheduling change the on-slot
fraction and slot rate and must be handled separately.

\hypertarget{methods}{%
\section{3. Methods}\label{methods}}

\hypertarget{published-experimental-benchmark}{%
\subsection{3.1 Published experimental
benchmark}\label{published-experimental-benchmark}}

Stancil \emph{et al.} report an average of approximately \(0.81\)
selected muon events for a beam-on pulse in the reduced-intensity
communication run {[}1{]}. Their measurement includes charged-current
interactions in upstream rock whose muons enter MINERvA, as well as a
smaller in-detector component. Consequently, it is incorrect to turn
that count directly into a fiducial-target interaction probability for
the detector alone.

As a check of Eq. (2), we assume independent pulses and zero background,
and pool \(n\) pulses for each bit, giving \(s = 0.81 n\). A fixed-seed
simulation of 200,000 equiprobable bits draws misses from
\(\exp(-0.81n)\). Its agreement with Eq. (2) checks the implementation
\textbf{only}: both use the same assumed probability, and neither
independently validates the experimental data. Stancil \emph{et al.}
report roughly 78\% correctly read uncoded bits in an individual
synchronized frame, about 99\% with five pooled frames, and no observed
errors for nine or more pooled frames; they also report performance
after convolutional decoding {[}1{]}. These are the published
observations against which the predicted bit-error \emph{scale} is
compared, without claiming a digitized measurement or an independent
dataset. Their published 0.1 bit s⁻¹ decoder-rate estimate is based on
their specific frame structure and a limited number of reused frame
combinations {[}1{]}.

For a \textbf{simulation calibrated to one measured geometry}, use the
published reduced-intensity NuMI run: \(2.25 \times 10^{13}\) 120 GeV
protons per on pulse and an estimated mean \({\lambda} = 0.81\) selected
muons per on pulse {[}1{]}. The paper estimates this mean as
\(2 \times 1402/3454\), assuming half the recorded slots are on.
Treating the 1,402 events as Poisson gives \({\hat{\lambda}} = 0.8118\)
with an approximate 95\% counting interval
\texttt{{[}0.7693,\ 0.8543{]}}; this covers input-counting uncertainty
only, not selection systematics. The on-pulse proton energy to the
target is about \texttt{0.433\ MJ}. This is neither facility electrical
consumption nor energy delivered as neutrinos. Nor can the 0.81 observed
muons be scaled linearly with an arbitrary receiver mass: upstream-rock
interactions contribute most of that sample {[}1{]}.

The accelerator timing is represented explicitly. The published schedule
has 25 pulse slots separated by 2.2 s, followed by a 6.267 s interval;
our schedule sets the 25 pulse times at offsets 0, 2.2, \ldots, 52.8 s,
and the next supercycle begins at 61.267 s {[}1{]}. Packet issue time is
uniform over the supercycle. Table 1b therefore reports mean and range
of modeled wall-clock latency over start phase, rather than multiplying
slot count by a mean interval. It assumes the stated repeated schedule
continues while a packet is sent.

Our protocol simulation carries five uniformly random payload bytes and
a CRC-8/ATM check byte (polynomial 0x07, zero initial state) with \(n\)
repeated on--off slots per bit: \(48n\) scheduled accelerator slots per
packet. Receivers know the packet boundary and slot phase externally.
Each on pulse produces independent Poisson counts with mean
\({\lambda}\); the off-slot mean is zero, approximating the experiment.
The decoder combines the \(n\) counts with a one-event threshold,
calculates the CRC and accepts only a matching check. We report
correctly accepted packets separately from all accepted packets and
CRC-undetected errors. At \(n = 3, 5, 9\), we simulate 50,000 random
messages at the central \({\hat{\lambda}}\) estimate and again at both
endpoints of its approximate 95\% counting interval. The central result
has separate finite-Monte-Carlo uncertainty; the endpoint runs provide
an input-mean sensitivity interval, not a combined confidence interval.
Synchronization is supplied externally, so acquisition and preamble
costs are absent. This is not a reproduction of the 2012 transmitted
40-bit word, 92 coded bits, 64-bit synchronization sequence or
convolutional decoder {[}1{]}. Inputs and outcomes are recorded by
\texttt{empirical\_message.py} and \texttt{empirical\_messages.csv}
{[}11{]}.

\hypertarget{illustrative-far-field-receiver-model}{%
\subsection{3.2 Illustrative far-field receiver
model}\label{illustrative-far-field-receiver-model}}

For a separate link sensitivity study, we assume an approximately
circular far-field neutrino footprint with characteristic half-angle
\({\theta}\). At distance \(L\), its area is
\(A = {\pi}({\theta}L)^{2}\). If the footprint is appreciably larger
than the receiver and illumination is sufficiently uniform, an
approximate selected-event probability per neutrino directed into that
footprint is

\[
p_sel ≃ (M/m_u) σ_CC(E) ε / [π(θL)^{2}] \tag{4}
\]

where \(M\) is fiducial target mass, \(m_u\) is the atomic mass unit
expressed in grams, \({\sigma}_CC\) is the per-nucleon charged-current
cross section and \({\varepsilon}\) subsumes trigger and event-selection
efficiency. This expression assumes a target small compared with the
beam footprint and does not specify a receiver shape. It cannot be
extrapolated to an arbitrarily narrow beam or a detector larger than the
footprint.

We set \(E = 3 GeV\),
\({\sigma}_CC = 0.67 \times 10^{-38}(E/GeV) cm^{2}\) per nucleon as an
approximate illustrative parametrization, \({\varepsilon} = 0.5\), and
vary \(M = 1, 10, 40 kt\), \({\theta} = 0.1, 1, 10 mrad\), and
\(L = 1,000, 5,000, 12,000 km\). The approximation gives
\(2.01 \times 10^{-38} cm^{2}\) per nucleon at 3 GeV. It is not a
precision calculation: T2K reports an inclusive charged-current
measurement on iron of \(2.29 ± 0.45 \times 10^{-38} cm^{2}\) per
nucleon at 3.3 GeV {[}28{]}, illustrating the need for a
detector-target-specific cross section and uncertainty {[}7{]}. We
assume a single selected flavor at the receiver, no oscillation loss, no
absorption, no off-slot background, perfect pointing and timing, and an
unconstrained beam of the specified divergence. Neither 1 mrad nor 0.1
mrad at 3 GeV has been established here for a source with the modelled
intensity. These are \textbf{free geometry parameters}, not a
demonstrated accelerator beam.

For fixed slot duration \({\tau}\) and slot rate \(R_slot = 1/{\tau}\),
the peak neutrino-carried power needed for an on slot is

\[
P_ν,on = R_slot × s_required × E / p_sel \tag{5}
\]

This is peak power during on slots (equivalently the full-duty average
if every slot is on), in watts when \(E\) is in joules. For equiprobable
uncoded OOK, the time-average is half this value. Equation (5) does not
compute the intensity or electrical power required to \emph{produce} and
\emph{focus} those neutrinos. For physically achievable sources, total
input power will depend on source conversion, energy distribution, duty
cycle and beam optics.

The numerical calculation uses \(m_u = 1.66053906660 \times 10^{-24} g\)
and \(1 GeV = 1.602176634 \times 10^{-10} J\). It is implemented in the
accompanying \texttt{model.py}, with fixed random seed \(20260923\); the
scenario grid and benchmark are supplied as CSV files {[}11{]}. The
finite-message example uses the \textbf{published NuMI source parameters
at one measured receiver}, whereas Eqs. (4)--(5) posit a different,
idealized monoenergetic far-field source. Their event rates and powers
cannot be joined by rescaling baseline alone.

\hypertarget{results}{%
\section{4. Results}\label{results}}

\hypertarget{communication-benchmark}{%
\subsection{4.1 Communication benchmark}\label{communication-benchmark}}

\begin{longtable}[]{@{}
  >{\raggedleft\arraybackslash}p{(\columnwidth - 6\tabcolsep) * \real{0.2500}}
  >{\raggedleft\arraybackslash}p{(\columnwidth - 6\tabcolsep) * \real{0.2500}}
  >{\raggedleft\arraybackslash}p{(\columnwidth - 6\tabcolsep) * \real{0.2500}}
  >{\raggedleft\arraybackslash}p{(\columnwidth - 6\tabcolsep) * \real{0.2500}}@{}}
\toprule\noalign{}
\begin{minipage}[b]{\linewidth}\raggedleft
Pulses pooled per raw bit
\end{minipage} & \begin{minipage}[b]{\linewidth}\raggedleft
Mean selected on-bit events
\end{minipage} & \begin{minipage}[b]{\linewidth}\raggedleft
Eq. (2) BER
\end{minipage} & \begin{minipage}[b]{\linewidth}\raggedleft
Simulated BER, 200,000 bits
\end{minipage} \\
\midrule\noalign{}
\endhead
\bottomrule\noalign{}
\endlastfoot
1 & 0.81 & 0.22243 & 0.22318 \\
5 & 4.05 & 0.00871 & 0.00893 \\
9 & 7.29 & 0.000341 & 0.000355 \\
\end{longtable}

\textbf{Table 1.} Analytic prediction and implementation check for
independently pooled on-pulse opportunities under ideal synchronization
and zero background. This table uses Stancil et al.'s published rounded
mean, \({\lambda} = 0.81\); Table 1b uses the event-count estimate
\({\hat{\lambda}} = 2 \times 1402/3454 = 0.8118\).

\begin{figure}
\centering
\includegraphics{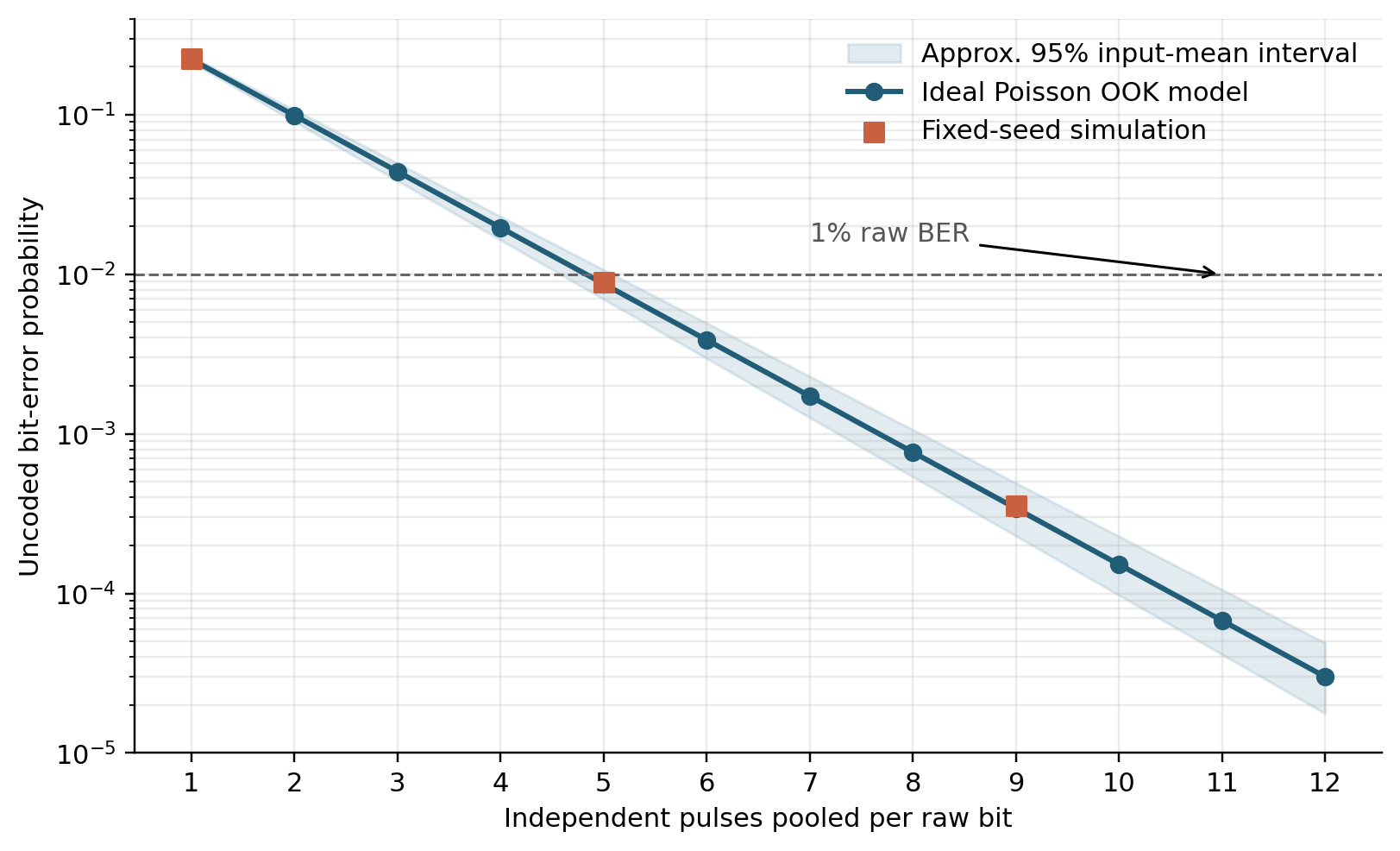}
\caption{Analytical raw bit-error probability and fixed-seed simulation
versus independently pooled pulses. Idealized zero-background,
synchronized on--off keying at the published selected-event mean, 0.81
per on pulse. The seeded points are drawn from the analytical model
itself and are an implementation check, not independent evidence for
it.}
\end{figure}

The simulation agrees with the analytical Bernoulli zero-count
prediction by construction. Five pooled on-pulse opportunities give
approximately 99.1\% correct \emph{uncoded synchronized bits} in this
model. Stancil \emph{et al.} report about 99\% correct bits from five
pooled \textbf{frames}; since each frame supplies a corresponding bit
opportunity, the count model agrees in scale, but their complete frame
synchronization, observed sample and convolutional decoder are not
reproduced here {[}1{]}.

Table 1b instead asks whether a short \emph{simulated} message survives
repeated bits and a CRC under the \textbf{published experimental
source--receiver count model}. The input mean
\({\hat{\lambda}} = 0.8118\) has approximate 95\% Poisson counting
limits of 0.7693--0.8543. All rows use 40 payload bits plus an eight-bit
CRC, external synchronization, 50,000 random messages per run and the
published pulse schedule. ``Correct accepted'' excludes both
CRC-rejected packets and wrongly accepted payloads. At the central mean,
the 95\% Monte Carlo margin for the correct-acceptance fraction near
0.66 is approximately ±0.004. The low/high input-mean rows in CSV
propagate counting uncertainty by simulation at the interval endpoints;
they do not include systematics or constitute a joint interval.

\begin{longtable}[]{@{}
  >{\raggedleft\arraybackslash}p{(\columnwidth - 12\tabcolsep) * \real{0.1429}}
  >{\raggedleft\arraybackslash}p{(\columnwidth - 12\tabcolsep) * \real{0.1429}}
  >{\raggedleft\arraybackslash}p{(\columnwidth - 12\tabcolsep) * \real{0.1429}}
  >{\raggedleft\arraybackslash}p{(\columnwidth - 12\tabcolsep) * \real{0.1429}}
  >{\raggedleft\arraybackslash}p{(\columnwidth - 12\tabcolsep) * \real{0.1429}}
  >{\raggedleft\arraybackslash}p{(\columnwidth - 12\tabcolsep) * \real{0.1429}}
  >{\raggedleft\arraybackslash}p{(\columnwidth - 12\tabcolsep) * \real{0.1429}}@{}}
\toprule\noalign{}
\begin{minipage}[b]{\linewidth}\raggedleft
Repeated on slots per bit
\end{minipage} & \begin{minipage}[b]{\linewidth}\raggedleft
Ideal raw bit BER
\end{minipage} & \begin{minipage}[b]{\linewidth}\raggedleft
Correct packets accepted
\end{minipage} & \begin{minipage}[b]{\linewidth}\raggedleft
Wrong payloads accepted
\end{minipage} & \begin{minipage}[b]{\linewidth}\raggedleft
Mean latency (min--max, s)
\end{minipage} & \begin{minipage}[b]{\linewidth}\raggedleft
Correct accepted payload bit/s
\end{minipage} & \begin{minipage}[b]{\linewidth}\raggedleft
Mean proton beam energy to target per attempt
\end{minipage} \\
\midrule\noalign{}
\endhead
\bottomrule\noalign{}
\endlastfoot
3 & 0.0438 & 11.47\% & 0.090\% & 351.5 (345.9--354.4) & 0.0130 & 31.1
MJ \\
5 & 0.00863 & 65.68\% & 0.002\% (1 of 50,000) & 586.9 (582.2--590.7) &
0.0448 & 51.8 MJ \\
9 & 0.000336 & 98.46\% & 0 observed in 50,000 & 1,057.6
(1,054.7--1,063.2) & 0.0372 & 93.3 MJ \\
\end{longtable}

\textbf{Table 1b.} Simulation calibrated to the published NuMI
selected-event mean, from \texttt{empirical\_messages.csv} {[}11{]}. The
reported latency range is over the possible starting phases of the
modeled supercycle for uniformly timed packet arrivals. ``Zero
observed'' does not establish zero true undetected-error probability.
The CRC detects errors; it does not correct them. The proton-energy
column counts on-pulse energy sent to the target, averaged over
messages; it excludes facility electricity. Rates exclude acquisition
and return acknowledgements and cannot be extrapolated to another
baseline with the far-field model.

At the lower and upper input-mean limits, simulated correct-acceptance
fractions are 8.8--15.3\% for three repeats, 60.0--71.9\% for five, and
97.65--98.92\% for nine. The corresponding accepted-correct payload
rates at five repeats are 0.0409--0.0490 bit/s. These intervals show
sensitivity to the estimated mean; their width also contains endpoint
Monte Carlo noise, and they omit systematics.

At five repetitions, the simulated on-pulse proton energy to the target
is 51.8 MJ per attempted packet, or about \textbf{2.0 MJ per correctly
accepted information bit} after accounting for rejected packets. These
numbers neither include power drawn by the accelerator complex nor
estimate energy carried by useful neutrinos. A one-packet command has a
mean modeled latency of about 587 s in this protocol, with approximately
582--591 s across start phase.

\hypertarget{illustrative-long-baseline-sensitivity}{%
\subsection{4.2 Illustrative long-baseline
sensitivity}\label{illustrative-long-baseline-sensitivity}}

Table 2 gives peak on-slot neutrino-carried power calculated from Eqs.
(3)--(5), for one-second slots (\(R_slot = 1\) slot/s) and
\({\theta} = 1 mrad\). All scenarios target 1\% uncoded bit error with
zero background, \textbf{unit flavor survival} and the same unverified
source divergence. For equiprobable uncoded OOK, time-average beam power
is half the table entry. The values should not be interpreted as
achievable source powers or physical lower bounds once their fixed
assumptions are relaxed.

\begin{longtable}[]{@{}rrrr@{}}
\toprule\noalign{}
Chord length & 1 kt receiver & 10 kt receiver & 40 kt receiver \\
\midrule\noalign{}
\endhead
\bottomrule\noalign{}
\endlastfoot
1,000 km & 9.76 MW & 0.976 MW & 0.244 MW \\
5,000 km & 244 MW & 24.4 MW & 6.10 MW \\
12,000 km & 1,405 MW & 140.5 MW & 35.1 MW \\
\end{longtable}

\begin{figure}
\centering
\includegraphics{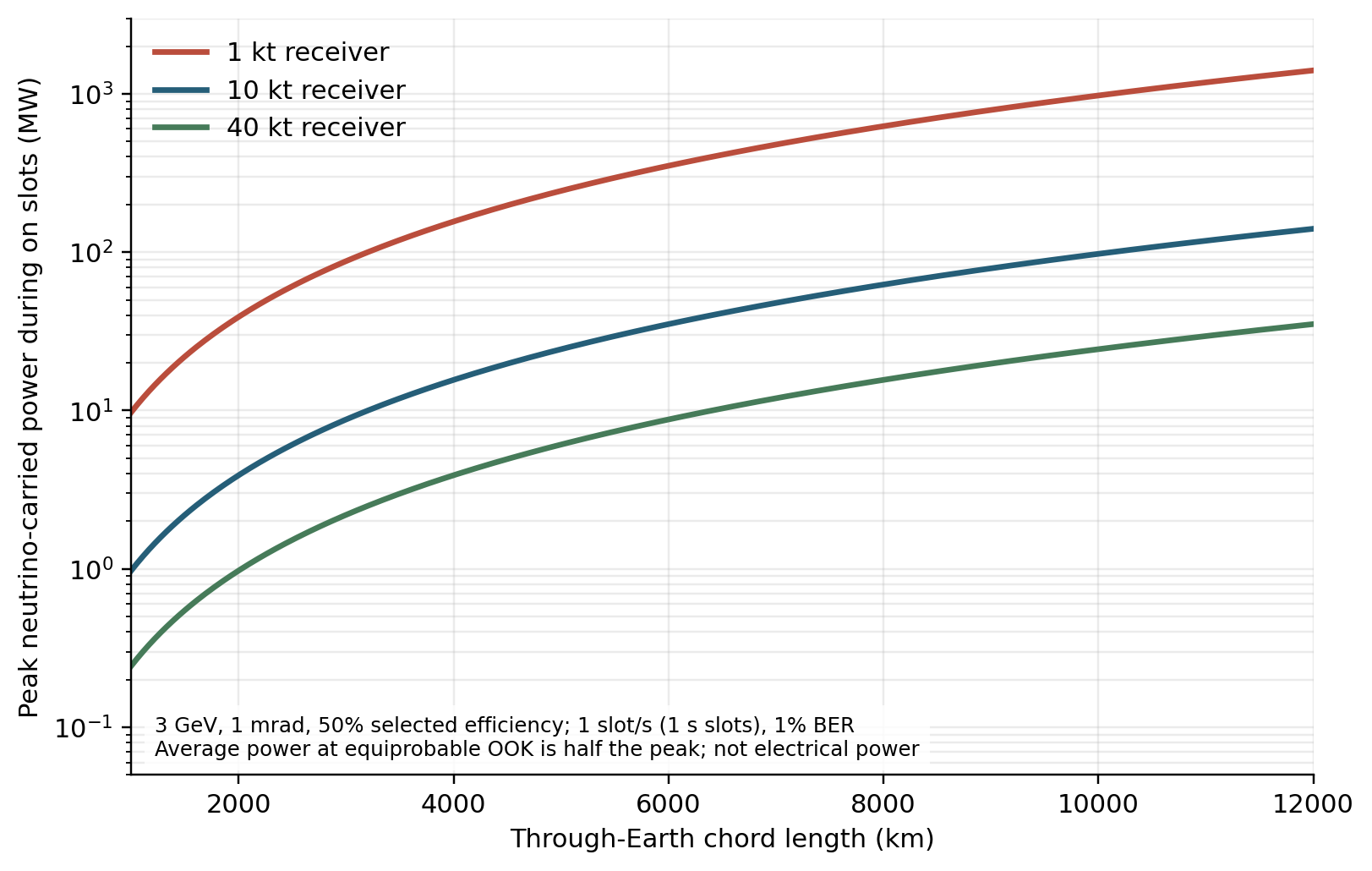}
\caption{Conditional peak on-slot neutrino-carried power versus chord
length for three detector masses. Conditional peak power while
transmitting an on slot; the illustrative slot rate is 1 slot/s
(one-second slots). For random equiprobable OOK, average
neutrino-carried power is half the ordinate. It is not accelerator input
power.}
\end{figure}

\begin{figure}
\centering
\includegraphics{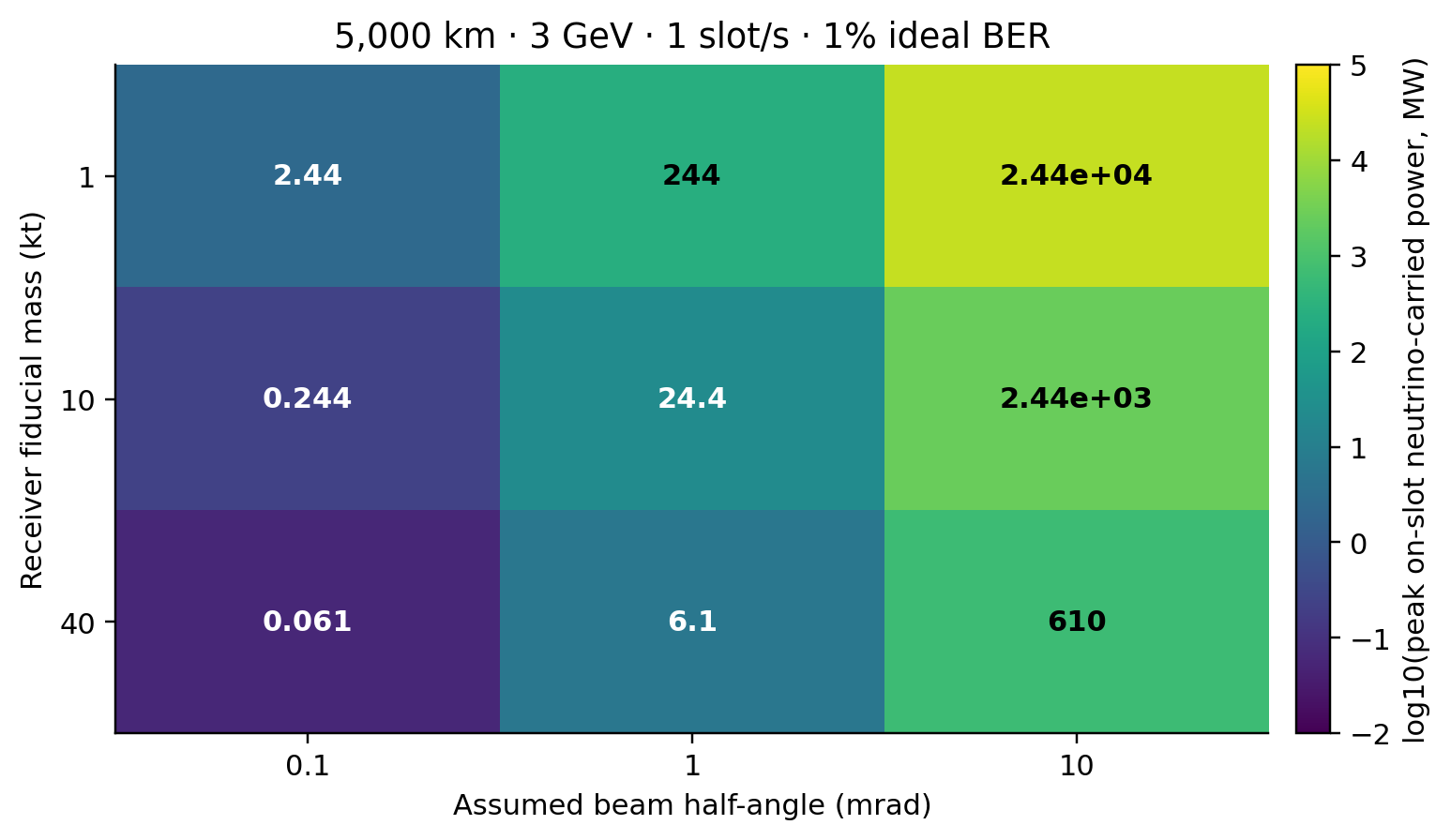}
\caption{Neutrino-carried energy-rate sensitivity to receiver mass and
assumed divergence at 5,000 km. This matrix visualizes nine of the 27
computed peak on-slot power cases at one slot/s. The 0.1 mrad column is
a mathematical sensitivity test; no source capable of that angular
spread at 3 GeV has been demonstrated here.}
\end{figure}

At fixed target mass and divergence, Eq. (4) gives
\(P_{\nu},on \propto L^{2} {\theta}^{2}/(M{\tau})\). The 10 kt, 5,000
km, one-second-slot reference case is 24.4 MW peak power during an on
slot at 1 mrad; the corresponding time average for equiprobable OOK is
12.2 MW. Setting the hypothetical divergence to 0.1 mrad changes the
peak reference value to 0.244 MW, while 10 mrad changes it to 2,440 MW.
Across all 27 calculated combinations, the smallest and largest entries
are 0.00244 MW and approximately 140,549 MW, respectively. These
extremes primarily reveal the leverage and danger of treating beam
divergence as an unconstrained parameter. The complete grid is in
{[}11{]}.

\begin{figure}
\centering
\includegraphics{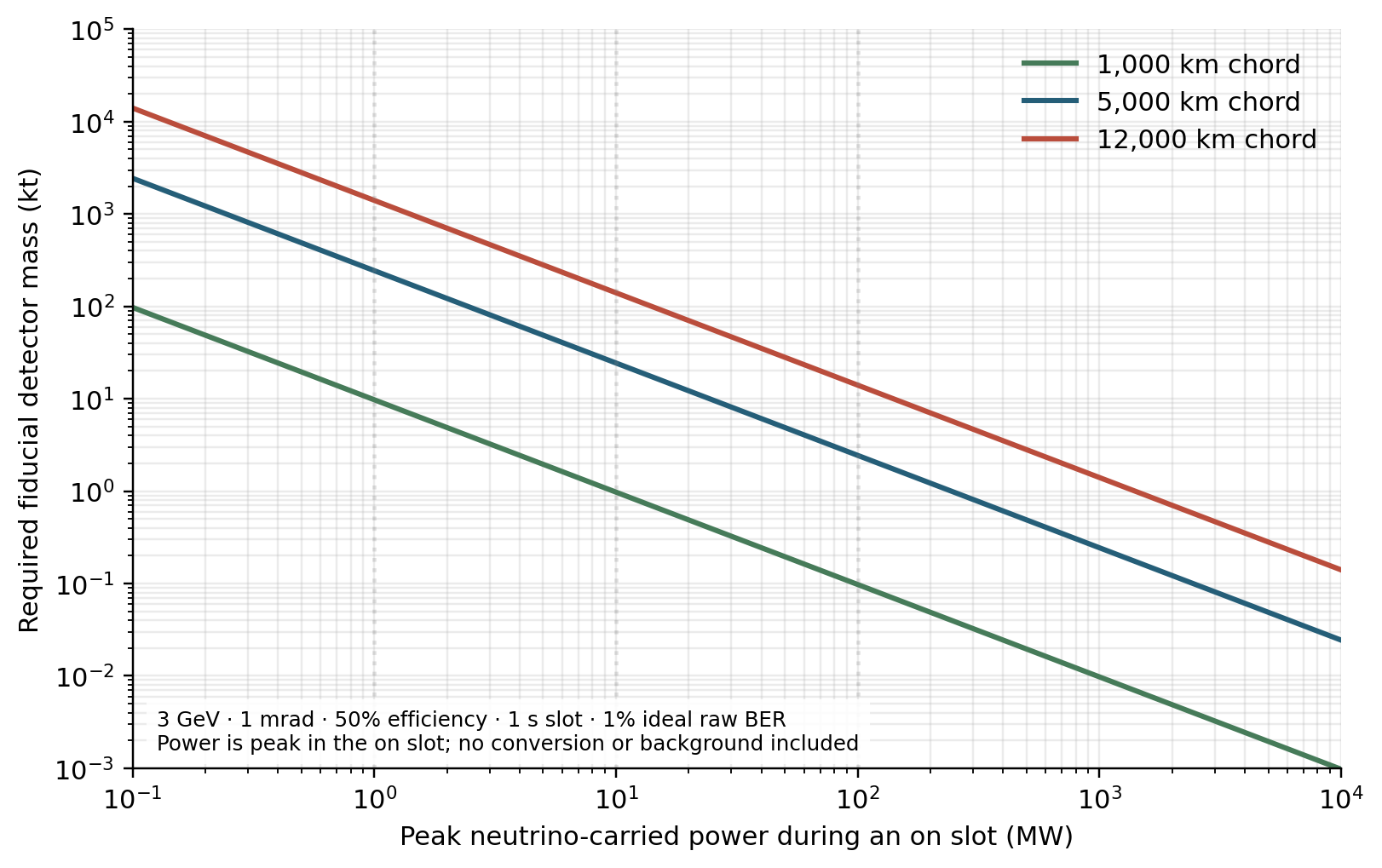}
\caption{Receiver mass required versus neutrino-carried on-slot power
for three chord lengths. Inversion of the same idealized model for
one-second slots (one slot/s). The ordinate is peak power during an on
slot. Each curve is a conditional mass--power trade-off, not a feasible
accelerator design.}
\end{figure}

The model yields a compact operating-point relation. At 1 mrad, 3 GeV
and 50\% efficiency, for a one-second on symbol and 1\% ideal raw BER,

\[
P_ν,on (MW) × M (kt) = 9.76 × (L/1,000 km)^{2} MW·kt \tag{6}
\]

More generally, the right-hand side multiplies by
\(({\theta}/1 mrad)^{2} \times (0.5/{\varepsilon}) \times (1 s/{\tau})\)
for slot duration \({\tau}\). Equation (6) assumes that the beam
footprint is wider than the detector and the other conditions of Section
3.2 hold. It states a \textbf{trade-off within one model}, not a minimum
cost of any possible communication system.

\begin{longtable}[]{@{}
  >{\raggedleft\arraybackslash}p{(\columnwidth - 8\tabcolsep) * \real{0.2000}}
  >{\raggedleft\arraybackslash}p{(\columnwidth - 8\tabcolsep) * \real{0.2000}}
  >{\raggedleft\arraybackslash}p{(\columnwidth - 8\tabcolsep) * \real{0.2000}}
  >{\raggedleft\arraybackslash}p{(\columnwidth - 8\tabcolsep) * \real{0.2000}}
  >{\raggedleft\arraybackslash}p{(\columnwidth - 8\tabcolsep) * \real{0.2000}}@{}}
\toprule\noalign{}
\begin{minipage}[b]{\linewidth}\raggedleft
Chord
\end{minipage} & \begin{minipage}[b]{\linewidth}\raggedleft
Product for 1 s on symbol
\end{minipage} & \begin{minipage}[b]{\linewidth}\raggedleft
Receiver at 1 MW carried by neutrinos
\end{minipage} & \begin{minipage}[b]{\linewidth}\raggedleft
Receiver at 10 MW
\end{minipage} & \begin{minipage}[b]{\linewidth}\raggedleft
Receiver at 100 MW
\end{minipage} \\
\midrule\noalign{}
\endhead
\bottomrule\noalign{}
\endlastfoot
1,000 km & 9.76 MW·kt & 9.76 kt & 0.976 kt & 0.0976 kt \\
5,000 km & 244 MW·kt & 244 kt & 24.4 kt & 2.44 kt \\
12,000 km & 1,405 MW·kt & 1,405 kt & 140.5 kt & 14.05 kt \\
\end{longtable}

These values are generated by \texttt{tradeoffs.py} and included in
\texttt{power\_detector\_tradeoff.csv} {[}11{]}. They expose the
receiver challenge: a 10 kt detector at 5,000 km requires 24.4 MW
carried by the assumed beam during a one-second on symbol; 100 MW
carried by neutrinos reduces the model receiver mass to 2.44 kt, still
far from a compact installation. Reducing the bit interval by a factor
of 1,000 raises the required neutrino-carried power by the same factor
at fixed mass. It cannot be compensated by faster neutrino propagation.

\textbf{Neutrino energy is distinct from source power.} In this
deliberately linear cross-section approximation
\({\sigma}_CC(E) \propto E\), the ratio \(E/{\sigma}_CC(E)\) in Eq. (5)
cancels when \({\theta}\) and \({\varepsilon}\) are artificially held
fixed. Raising the individual neutrino energy from 3 GeV does
\textbf{not}, by itself, improve this simplified sensitivity
calculation. A physical source changes its energy--angle spectrum and
attainable divergence with accelerator design and neutrino energy;
interaction cross sections, detector response, Earth propagation and
secondary-particle ranges also change. The proton or muon beam energy
and wall-plug power are separate quantities. If \(η_useful\) denotes the
total efficiency from facility input to energy carried by neutrinos in
the useful beam, the facility input would be
\(P_input = P_{\nu}/η_useful\), before any other site loads. We have not
established \(η_useful\) for a specified design and therefore cannot
give a credible accelerator electrical-power number. For scale only, the
published communication run's \textbf{all-on proton-pulse energy
averaged over its supercycle} is approximately 0.177 MW, whereas this
\emph{different source model} requires 24.4 MW carried by useful
neutrinos for the 5,000 km, 10 kt, one-slot/s reference. This is an
energy-budget comparison across distinct setups, not a way to
extrapolate NuMI's angular flux or selected event rate to 5,000 km.

The displayed values are conditional, optimistic estimates. They are
\textbf{not} a claim that the desired rates are feasible, nor a
universal lower bound on all possible neutrino communication systems. A
beam energy spectrum, realistic source geometry or another detection
channel changes the result; backgrounds, pointing losses, flavor
oscillations and synchronization generally increase the requirements of
this particular link design.

\hypertarget{raw-coded-and-compressed-throughput}{%
\subsection{4.3 Raw, coded and compressed
throughput}\label{raw-coded-and-compressed-throughput}}

Equation (7) can be inverted into a deliberately normalized \emph{raw
slot rate}, holding 3 GeV, 50\% efficiency, 1 mrad and 1\% ideal uncoded
bit error fixed:

\[
R_raw ≃ 0.1025 [P_ν,on/MW] [M/kt] (1,000 km/L)^{2} slot s^{-1} \tag{7}
\]

For another divergence or efficiency, multiply by
\((1 mrad/{\theta})^{2}({\varepsilon}/0.5)\). In ordinary mass units,
\(M/kt = (M/kg)/10^{6}\). This normalization is meaningful only for the
model's \textbf{neutrino-carried peak on-slot power}, receiver mass and
raw Poisson decision. \(R_raw\) counts every OOK slot, each carrying one
uncoded bit, including both zero and one slots; it is not a rate per
accelerator MW. For equiprobable OOK, time-average power is about
\(P{\nu},on/2\), while peak on-slot power is unchanged. Coding can
change slot rate and on-slot fraction.

To expose the difference between physical rate and useful rate, define a
\emph{bookkeeping example} with forward-error-correction (FEC) code rate
\(r_c = 1/2\) and a frame payload fraction \(f = 0.8\). Then
\(R_payload,illustrative = R_raw r_c f = 0.4 R_raw\), before
acquisition, retransmission, acknowledgements or decoder failures. No
specific code has been shown to achieve a target post-decoder error rate
under this neutrino count distribution; the 1\% target refers only to
the \textbf{uncoded} count decision. Coding can also allow operation at
a higher raw error rate and reduce required events per symbol, so a
fully optimized coded link cannot be estimated by multiplying one
uncoded operating point by 0.4.

For a 5,000 km chord and 10 kt target under these assumptions, the power
column is peak on-slot power. A 100 MW entry corresponds to a 4.10
slot/s model rate, or \({\tau} \approx 0.244 s\) slots; at equiprobable
uncoded OOK its time-average power would be about 50 MW.

\begin{longtable}[]{@{}
  >{\raggedleft\arraybackslash}p{(\columnwidth - 6\tabcolsep) * \real{0.2500}}
  >{\raggedleft\arraybackslash}p{(\columnwidth - 6\tabcolsep) * \real{0.2500}}
  >{\raggedleft\arraybackslash}p{(\columnwidth - 6\tabcolsep) * \real{0.2500}}
  >{\raggedleft\arraybackslash}p{(\columnwidth - 6\tabcolsep) * \real{0.2500}}@{}}
\toprule\noalign{}
\begin{minipage}[b]{\linewidth}\raggedleft
Peak on-slot neutrino power
\end{minipage} & \begin{minipage}[b]{\linewidth}\raggedleft
Raw slots/s at 1\% uncoded BER
\end{minipage} & \begin{minipage}[b]{\linewidth}\raggedleft
Bookkeeping payload bits/s (assumed factors; no decoder)
\end{minipage} & \begin{minipage}[b]{\linewidth}\raggedleft
If a long telemetry stream compresses 4:1:
uncompressed-source-equivalent bits/s
\end{minipage} \\
\midrule\noalign{}
\endhead
\bottomrule\noalign{}
\endlastfoot
1 MW & 0.0410 & 0.0164 & 0.0656 \\
10 MW & 0.410 & 0.164 & 0.656 \\
100 MW & \textbf{4.10} & \textbf{1.64} & \textbf{6.56} \\
\end{longtable}

The rightmost column is \textbf{not} a faster physical link. The
illustrative bookkeeping payload remains 1.64 bit s⁻¹ in the 100 MW row;
a hypothetical 4:1 lossless compressor would merely represent a source
stream that previously used 6.56 bit s⁻¹. The compression factor is an
explicit optimistic assumption for repetitive telemetry, not a measured
result and not a plausible default for a tiny already encoded market
order. Framing fractions also deteriorate for short messages. The 100 MW
example assumes a large fixed detector and an unestablished 100 MW of
useful neutrino-carried on-slot power.

A compact figure of merit for the \textbf{assumed coded example} is

\[
Q = R_payload L^{2}/(P_ν,on M) ≃ 0.0410 bit·km^{2}/(s·MW·kg) \tag{8}
\]

with \(L\) in km and \(M\) in kg. Equivalently, at 5,000 km the model
yields approximately \(0.00164 payload bit s^{-1}/(MW·kt)\), or
\(1.64\times10^{-9} payload bit s^{-1}/(MW·kg)\). These normalized
numbers must always be quoted with the energy, divergence, selection
efficiency, raw BER, assumed code rate and framing fraction.
\texttt{throughput.py} generates the 27-case
\texttt{illustrative\_throughput.csv} {[}11{]}.

\begin{figure}
\centering
\includegraphics{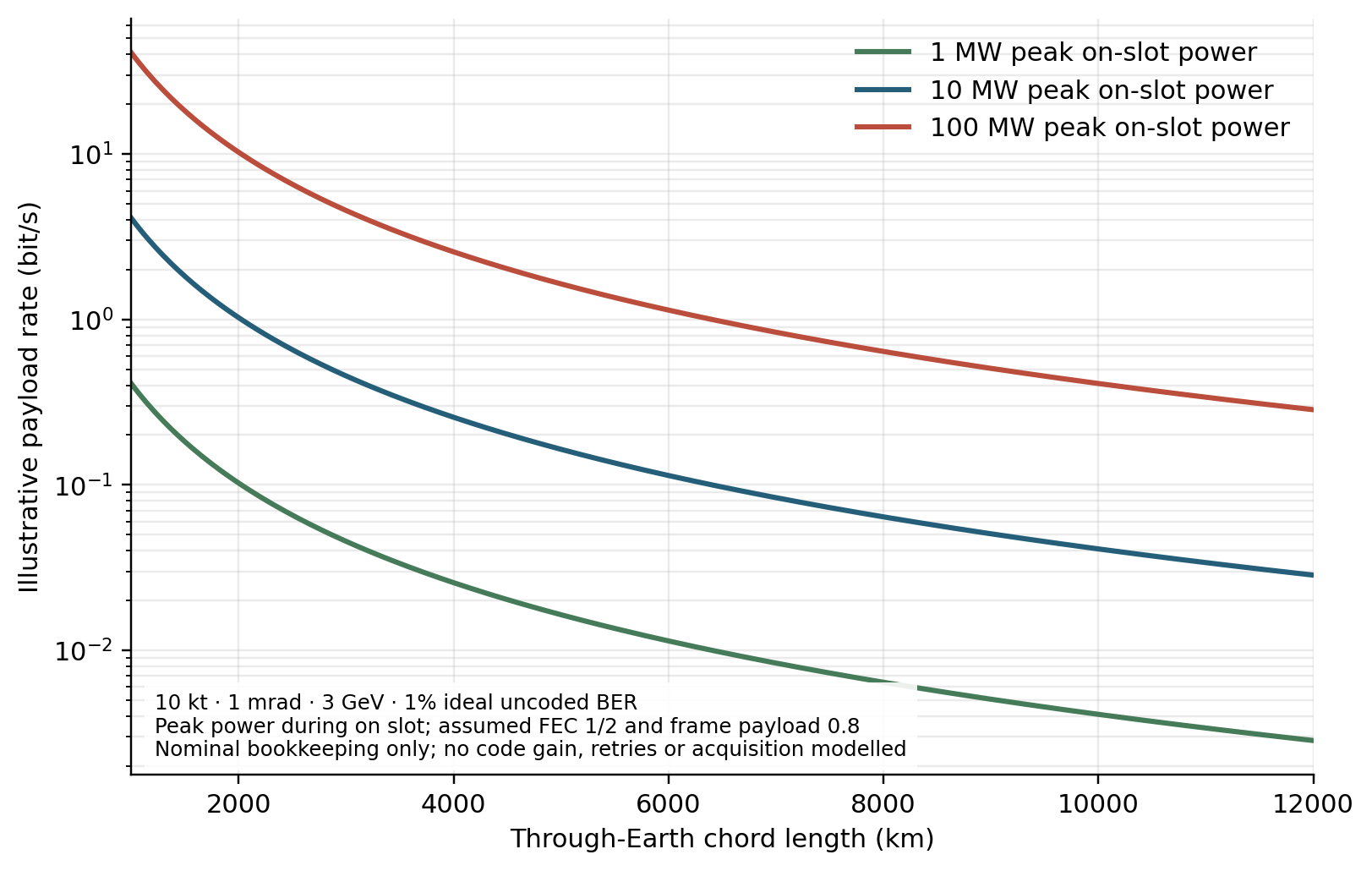}
\caption{Nominal bookkeeping payload rate versus chord for three assumed
peak on-slot power levels. Nominal bookkeeping illustration with 10 kt
mass, half-rate FEC and 80\% frame payload fraction. The curves assume
the same 1\% uncoded event threshold and do not measure performance of
an implemented decoder. Power is peak during on slots; the plotted rate
counts all raw OOK slots.}
\end{figure}

\hypertarget{discussion}{%
\section{5. Discussion}\label{discussion}}

\hypertarget{relation-to-earlier-work}{%
\subsection{5.1 Relation to earlier
work}\label{relation-to-earlier-work}}

The Fermilab demonstration establishes message transfer and supplies a
measured count distribution against which a simple Poisson decoder can
be checked {[}1{]}. Huber's submarine analysis presents a specific
proposed source and receiver concept {[}2{]}. The present long-baseline
sweep does not supersede either result: it supplies a common, explicitly
simplified accounting of how detected counts constrain communication and
how one assumed geometry scales. The finite-packet exercise uses Stancil
\emph{et al.}'s measured source--receiver combination, but its CRC and
repetition protocol is our construction; it does not predict event rates
for other paths.

The relationship to {[}5{]} is methodological. Both studies require the
analyst to separate a statistical detection score from a concrete
operating point and to make detector geometry explicit. Passive
submarine detection in {[}5{]} and active communication here have
different energies, source control, backgrounds and receiver objectives.
Direct citation to {[}5{]} should introduce this change of question and
any legitimately reused methods, rather than imply that its antineutrino
sensitivity numbers validate Eq. (5). Hallsjö's detector thesis {[}6{]}
likewise motivates scrutiny of selection efficiency and interaction
reconstruction but is not evidence that a 50\% selected efficiency
applies to the receiver postulated here.

\hypertarget{low-rate-communication-strategies-from-deep-space-research}{%
\subsection{5.2 Low-rate communication strategies from deep-space
research}\label{low-rate-communication-strategies-from-deep-space-research}}

A useful analogue comes from NASA/JPL's response to Galileo's failed
high-gain antenna. Statman describes a revised low-gain link combining
compression, arraying of ground antennas, convolutional and
variable-redundancy Reed--Solomon coding, decoding feedback and
reprocessing of recorded data {[}13{]}. These are system-level
techniques for recovering useful information from a weak link. Their
numerical radio-link gains cannot be transferred to a neutrino beam, but
they sharpen the question this paper should ask: how much \emph{verified
payload} arrives per unit time and source energy after acquisition and
error control?

Moision and Hamkins analyze an optical photon-counting deep-space
channel, jointly selecting pulse-position modulation (PPM) order and
error-control code rate under average and peak power constraints
{[}14{]}. Poisson event counting is the useful mathematical connection
to Eq. (1). In the neutrino case, pulse timing, allowed accelerator
patterns, beam-on energy, background and receiver dead time determine
whether PPM or another sparse-pulse code can outperform simple on--off
keying. No such gain is assumed in Tables 1--2. NASA/JPL's work on joint
decoder and frame synchronization at extremely low data rates also
supports treating symbol acquisition as a measured part of the link
rather than granting it free of charge {[}15{]}.

Stancil \emph{et al.} give the zero-background OOK Poisson-channel
capacity
\(C/R_{\mathrm{slot}} = \log_2[1 + (1 - e^{-\lambda})\exp(-\lambda/(e^{\lambda} - 1))]\),
approximately \textbf{0.37 bits per pulse} at \(\lambda \approx 0.81\)
{[}1{]}. Our five-repeat CRC simulation yields about
\(40 \times 0.6568/(48 \times 5) \approx 0.109\) correctly accepted
payload bits per scheduled slot at the central input mean. These are not
directly equivalent performance measurements: 0.109 is a finite-packet
payload yield per scheduled slot for one assumed protocol, while 0.37 is
an asymptotic channel-capacity reference under Stancil et al.'s stated
channel model. Neither value establishes an optimized code or the
delivered rate of a practical link.

For the next analysis, compare OOK, repetition with soft count
combining, and one constrained PPM scheme using the \textbf{same}
average source-energy budget and an explicit maximum pulse rate. For
each scheme report detection, false acquisition, coding and framing
overhead; packet success probability; and delivered information bits per
second and per joule. Compression should be evaluated only on a
specified source message distribution; random or already compressed
payloads do not offer a free gain. Combining counts from multiple
receivers is a possible analogue of antenna arraying only after their
acceptance, timing and independence are modelled. Delay-tolerant
networking may help carry an intermittent link's data end to end, but it
does not increase the underlying detected-event rate and is outside this
one-hop calculation.

\hypertarget{coding-and-source-compression-choices}{%
\subsection{5.3 Coding and source compression
choices}\label{coding-and-source-compression-choices}}

Galileo's low-rate recovery used compression and concatenated
error-control codes {[}13{]}. For the present link, compare repetition
and soft count combining against a finite-blocklength convolutional or
LDPC code and a constrained sparse-pulse scheme, with synchronization
and energy budgets matched. The CCSDS telemetry synchronization and
channel coding standard {[}21{]} supplies established design families,
not measured coding gains for neutrino counts. Decoder latency and
finite frame length are critical in financial messages; a powerful
long-block code may deliver fewer bits before a deadline than a short
simpler code.

Lossless source coding should be chosen for the message distribution.
CCSDS 121.0-B-3 specifies a lossless telemetry compression method
{[}20{]}; Zstandard with a pre-shared dictionary is another documented
general-purpose lossless option {[}22{]}. A fixed dictionary of
permitted short commands can use fewer transmitted bits than verbose
text. Compressors can expand short or nearly random messages after
headers; demonstrate a 4:1 factor on an actual representative corpus
before using the rightmost column of the throughput table in Section
4.3. Lossy compression may suit some imagery but cannot be applied to
exact commands, orders or scientific measurements without an
application-specific error budget. Neither source compression nor FEC
changes neutrino interaction probability: compression removes source
redundancy, while FEC adds channel redundancy to improve recoverability.

\hypertarget{submerged-receivers-and-submarine-communication}{%
\subsection{5.4 Submerged receivers and submarine
communication}\label{submerged-receivers-and-submarine-communication}}

The submerged-receiver application has a different objective and
constraint set from the fixed 10 kt reference. Huber's earlier proposal
{[}2{]} explicitly considers a one-way high-energy beam from a muon
storage ring and reception through muons produced either in the
submarine or in surrounding water. That can enlarge the effective
interaction volume beyond the on-board apparatus. Our mass-only, 3 GeV
charged-current target model does \textbf{not} include that mechanism
and cannot validate or exclude Huber's proposed rates.

It does show the price of simply miniaturizing the receiver while
keeping our fixed-beam assumptions. At a 1,000 km chord, a 100 tonne
fiducial target (\(0.1\,\mathrm{kt}\)) corresponds to about 97.6 MW
carried by neutrinos during a one-second on symbol; a 10 tonne target
(\(0.01\,\mathrm{kt}\)) corresponds to about 976 MW. At 5,000 km these
become approximately 2,440 and 24,400 MW, respectively. These are
extrapolations of Eq. (6), \textbf{not} numbers for Huber's
water-produced-muon receiver. A practical study must model the beam's
illuminated area in seawater, muon production and range, optical
background, submarine depth and motion, steering and location
uncertainty, plus receiver size and energy supply. A one-way downlink
can deliver short commands without an onboard accelerator, but
acknowledgement or a return data path would require a separate system. A
submarine's location also cannot be presumed known precisely enough to
keep a narrow beam aligned.

\hypertarget{low-latency-financial-communication-propagation-versus-decoding}{%
\subsection{5.5 Low-latency financial communication: propagation versus
decoding}\label{low-latency-financial-communication-propagation-versus-decoding}}

Short messages between distant financial centres are a plausible reason
to examine a direct chord. Research on low-latency trading emphasizes
the competition between message reliability and time to decode {[}17{]},
while measurements of Chicago--New York links document the value of
optimized fiber and near-line-of-sight microwave routes {[}18{]}.
Neither finding establishes a viable neutrino trading link.

Consider a \textbf{hypothetical 5,000 km Earth chord}, without naming
real trading venues. Neutrino flight time is about
\(L/c = 16.68\,\mathrm{ms}\). The shortest surface arc between the same
endpoints is about 5,138 km. Using \(n_g = 1.4677\) as an illustrative
group index of a specified conventional optical fiber {[}19{]}, ideal
propagation along that surface arc is \(n_g s/c = 25.15\,\mathrm{ms}\).
The maximal propagation-only advantage against this idealized
shortest-surface-fiber example is \textbf{8.48 ms one-way}. Actual cable
routing may lengthen the fiber path, while free-space or alternative
fiber technologies may offer much smaller differences. End-to-end
comparisons must include source scheduling, message acquisition,
decoding, routing and any acknowledgement.

\begin{figure}
\centering
\includegraphics{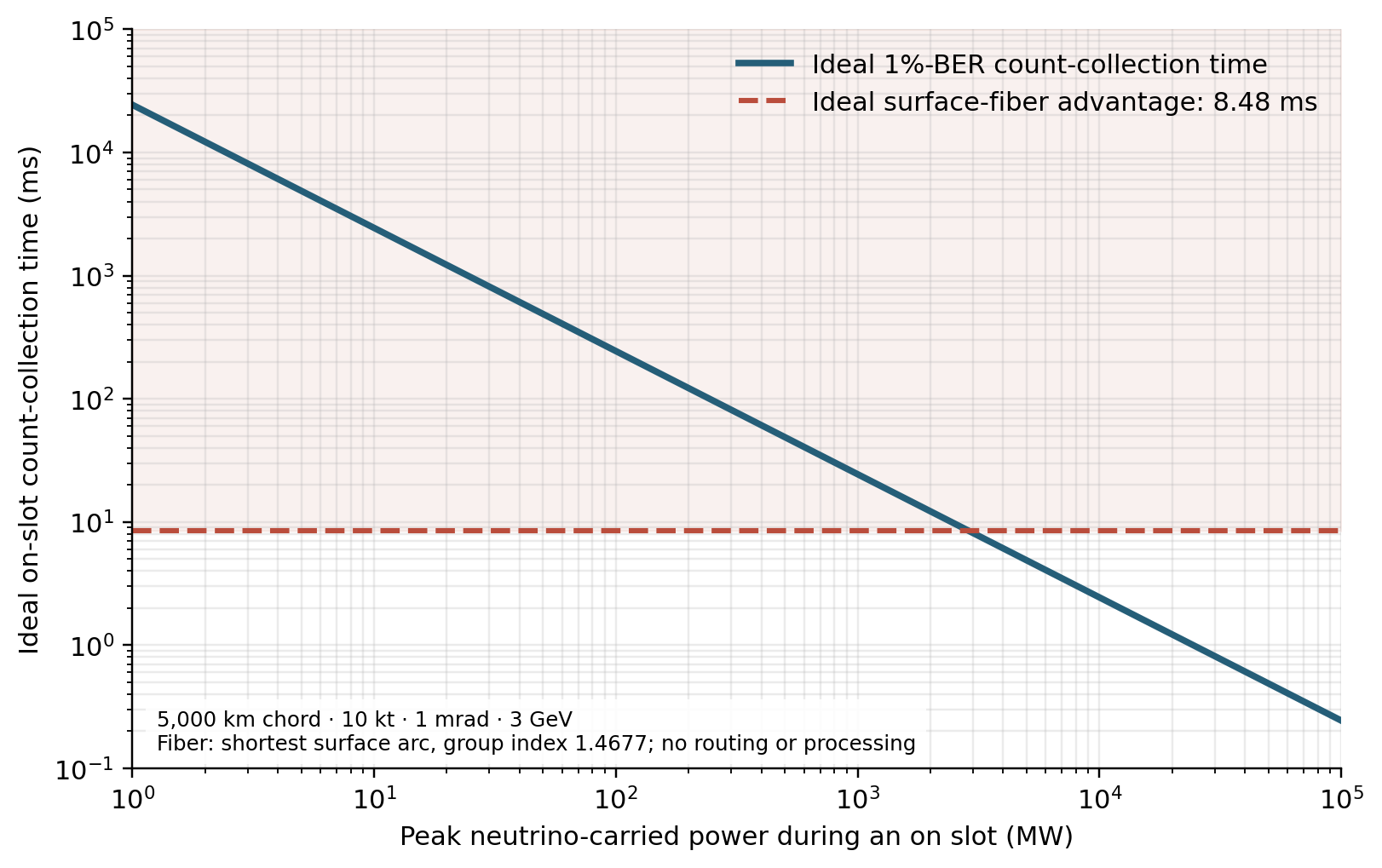}
\caption{Ideal count-collection time versus neutrino-carried power
compared with an illustrative fiber propagation advantage. The
horizontal line is a propagation-only comparison for one hypothetical
geometry. The sloping line is the minimum \emph{modelled on-slot
count-collection time} to reach 1\% raw BER with a 10 kt receiver.
Synchronization, packet overhead and processing would add delay.}
\end{figure}

In the same idealized 10 kt, 1 mrad receiver, 100 MW carried by
neutrinos requires about \(244\,\mathrm{ms}\) to collect the target
expected signal for one on symbol. Reaching \(1\,\mathrm{ms}\) would
require about \textbf{24.4 GW} carried by neutrinos; reaching the
\(8.48\,\mathrm{ms}\) propagation-only break-even window requires about
\textbf{2.88 GW}. At 100 MW, that break-even condition instead implies
an approximately \textbf{288 kt} receiver. These are conditional
inversions of Eq. (6) with a one-bit, 1\% raw-error criterion and unit
flavor survival; they are especially fragile to the omitted beam and
propagation physics. A transaction message, clock acquisition, coding,
confirmation and operational redundancy add time or error constraints.
The sign of the apparent speed advantage can therefore reverse once
decoding is included. A one-way trading signal is also not a complete
order-execution and acknowledgement path. The example identifies a
research question---whether any physically achievable
transmitter--receiver design could meet a specified end-to-end
deadline---rather than a positive business case.

\hypertarget{limits-of-the-first-calculation}{%
\subsection{5.6 Limits of the first
calculation}\label{limits-of-the-first-calculation}}

\textbf{Source feasibility.} The strongest unverified assumption is the
angular distribution of useful neutrinos at the selected energy. A
genuine accelerator study must link pion or stored-muon production,
focusing, decay geometry, proton or muon current, duty cycle, beam
losses, and resulting energy--angle correlations. The generated flux
must be integrated over the actual receiver footprint. Neutrino-carried
energy is only one component of that calculation. The nuSTORM study
{[}8{]} is a relevant source-design starting point.

\textbf{Propagation and interaction.} The calculation holds flavor
survival at one. To expose the size of this omission, a \emph{vacuum,
two-flavor illustration only} with
\(\Delta m^2 = 2.5 \times 10^{-3}\,\mathrm{eV}^2\) and maximal mixing
yields
\(P(\nu_\mu\to\nu_\mu) \approx 1 - \sin^2[1.267\,\Delta m^2(\mathrm{eV}^2)L(\mathrm{km})/E(\mathrm{GeV})] \approx 0.29\)
at 5,000 km and 3 GeV. If that illustrative factor simply multiplied
selected events, the corresponding power entry would grow by about
3.5-fold. It is \textbf{not} a terrestrial prediction: three-flavor
oscillations, matter profile, spectrum and flavor-sensitive efficiency
change the answer {[}9{]}. A narrow 3 GeV beam also does not represent
the published NuMI spectrum, which peaks near 3.2 GeV with a broad width
{[}1{]}. The appropriate rate is an energy and angle integral over
source flux, oscillation probability, Earth transmission,
detector-target-specific cross section and efficiency, with propagated
uncertainties {[}7,28{]}. At these energies absorption may be small
compared with oscillation effects for Earth chords, but must still be
checked for the chosen geometry and source.

\textbf{Detector and receiver geometry.} A mass-only model hides
projected area, target depth, channel choice, cosmic and atmospheric
backgrounds, event containment, trigger and timing. DUNE and
Hyper-Kamiokande designs {[}10,12{]} establish examples of large
stationary detector technologies, but neither design report demonstrates
a dedicated communication receiver. A mobile submarine receiver requires
its own volume, power, timing and mechanical analysis, and should not be
grouped with fixed underground detectors in an engineering claim.

\textbf{Information delivery.} Eq. (2) assumes known slot boundaries and
zero background. Table 1b now supplies one finite-message CRC and
repetition example with the \emph{published} NuMI pulse schedule, and
records both correct accepted and undetected erroneous packets. It still
assumes external synchronization, independent stationary event counts,
no beam losses and no acknowledgements. For a prospective source,
measure nonzero-background false-positive and false-negative
probabilities and simulate synchronization, framing, code decoding,
dropped slots and total message completion. The experimental 0.1 bit s⁻¹
decoded rate follows the actual published frame and convolutional code
{[}1{]}; it cannot be inferred from 0.81 events per pulse alone, nor
compared as a like-for-like rate with our different protocol. The
far-field \texttt{1\ raw\ slot\ s⁻¹} row remains a comparison
convention, not a payload-throughput prediction.

\textbf{Use-case comparison.} Any claim of practicality also requires a
specified use case and competing communication path. A fixed global
direct link, a submerged mobile receiver and a planetary occultation
link have different requirements. The value of direct propagation cannot
be inferred solely from the achievable event rate.

\hypertarget{work-required-for-a-submission-quality-article}{%
\section{6. Work required for a submission-quality
article}\label{work-required-for-a-submission-quality-article}}

The empirical NuMI analysis now uses its published proton intensity,
time structure and selected-count mean for a \emph{different}
finite-packet protocol. The next revision must (i) choose a specific
source design with a published \textbf{spectral angular flux at the
receiver} and source electrical or proton-beam power; (ii) propagate
that spectrum through a stated Earth density profile with three-flavor
oscillations and an energy-dependent interaction model; (iii) select a
receiver material, projected area and response curve, including
rock-produced secondaries if relevant; (iv) incorporate measured or
defensibly projected background in actual synchronized time windows; and
(v) compare complete coded messages at equal average source energy,
including a sparse-pulse scheme motivated by {[}13--15{]}, before
reporting delivered information per elapsed second and per joule of
facility input. A deeper reproduction of the 2012 experiment would need
its actual frame timing, source interruptions, convolutional decoder and
preferably the underlying observed counts; Table 1b is \textbf{not} that
reconstruction. Finally, use uncertainty ranges on flux, propagation,
cross section, selection and acquisition in any source--mass--rate
feasibility contour.

These are material missing analyses, not editorial refinements. Until
completed, the numerical power values in Table 2 should be presented
only as a sensitivity map.

\hypertarget{conclusion}{%
\section{7. Conclusion}\label{conclusion}}

The experimental anchor is Stancil \emph{et al.}'s demonstrated 1.035 km
link {[}1{]}. In our separate simulation calibrated to its
selected-event mean, five repeated slots and CRC correctly accept 65.7\%
of simulated packets carrying a 40-bit payload on average; modeled mean
packet latency is 587 s with a 582--591 s phase range. The estimated
Poisson uncertainty in the input mean shifts this acceptance to about
60--72\%. These are simulated outcomes under known synchronization, not
packet measurements or a recreation of the original decoder.

The independent far-field model remains a geometric sensitivity
calculation. Its headline 5,000 km result---\textbf{24.4 MW peak
neutrino-carried power during on slots} for a 10 kt receiver---is
conditional on 3 GeV monoenergetic neutrinos, 1 mrad divergence, 50\%
selected efficiency, unit flavor survival, zero background, and
one-second slots (1 slot/s). For equiprobable uncoded OOK, average power
is half the peak. No source--detector installation meeting these
assumptions has been established.

Three constraints dominate the interpretation. First, useful source
intensity and focusing must be physically demonstrated and converted to
facility power. Second, the detector must collect enough \emph{selected}
events against background within an actual decoding deadline, including
alignment and synchronization. Third, a short decoded bit must be
embedded in an end-to-end message protocol whose payload rate and
reliability can be measured. Changing the neutrino's energy does not
automatically evade the energy--interaction trade-off in our linear
cross-section approximation; the coupled spectrum and beam optics
require explicit modelling.

The separate long-stream example remains bookkeeping: 100 MW peak
on-slot power, 10 kt, 5,000 km, assumed half-rate FEC and 80\% frame
payload produce 4.10 raw slots/s and 1.64 nominal payload bits/s before
any decoder-performance or packet-success estimate. A hypothetical 4:1
compression factor changes only the represented source volume. Neither
figure is measured payload throughput. Submarine, finance and Solar
System cases remain scoped illustrations because each needs a distinct
source, receiver and propagation model.

Huber's submarine concept uses a water-assisted muon receiver and needs
a dedicated simulation {[}2{]}. In the financial illustration, an ideal
5,000 km chord has an 8.48 ms propagation-only advantage over the stated
shortest-surface conventional-fiber comparison; the reference model
needs about 2.88 GW peak neutrino-carried on-slot power with a 10 kt
receiver just to collect one symbol within that interval, before
protocol overhead. Neither application is shown to be practical. A
source-specific beam, detector, propagation and protocol model is
required for engineering or economic conclusions.

\hypertarget{appendix-a.-occulted-links-within-the-solar-system}{%
\section{Appendix A. Occulted links within the Solar
System}\label{appendix-a.-occulted-links-within-the-solar-system}}

Neutrino signalling could in principle reach a receiver behind
intervening matter when an electromagnetic path is obstructed; Stancil
\emph{et al.} explicitly mentioned planetary-body blockage as a possible
use {[}1{]}. Two distinct cases warrant examination. A lunar far-side
installation has no direct view of Earth because the Moon lies between
transmitter and receiver; NASA is developing lunar relay services for
precisely that coverage gap {[}23{]}. During Mars solar conjunction,
radio links are disrupted by the solar corona over an interval wider
than any literal occultation by the solar disk; NASA commonly limits
commanding then {[}24{]}. A direct neutrino path near or through the Sun
is a separate, much less studied alternative to a radio or optical relay
{[}25{]}. Neither case has an established neutrino source--receiver
design; the new material here is an explicit, conditional
distance-scaling calculation.

To put scale on the proposal, extend Eqs. (3)--(5) \emph{only as a
geometric sensitivity calculation}. Keep the hypothetical 3 GeV
single-flavor beam, 1 mrad far-field half-angle, 10 kt receiver, 50\%
selected-event efficiency, zero background and 1\% uncoded BER for a
one-second slot. With no intervening-body absorption or flavor
conversion and perfect alignment, peak neutrino-carried on-slot power is

\[
P_{\nu,\mathrm{on}} = 0.976\,\mathrm{MW} \times (L/1000\,\mathrm{km})^2 \times (10\,\mathrm{kt}/M) \times (\theta/1\,\mathrm{mrad})^2 \times (R_{\mathrm{slot}}/1\,\mathrm{slot\,s^{-1}}) \tag{A1}
\]

The corresponding selected-event target remains 3.912 per on slot. The
estimates below use one-second slots (one slot/s) and report peak
neutrino-carried on-slot power. For equiprobable OOK the time average is
half the listed values. The values are computed by
\texttt{communication/analysis/solar\_system\_scaling.py} using the same
function and constants as the terrestrial model; they are energy flow in
neutrinos, \textbf{not accelerator electrical power}. Distances use
NASA's mean Earth--Moon separation and the defined astronomical unit
{[}26,27{]}. A far-side surface link would be slightly longer than the
centre-to-centre lunar distance, so the lunar row is a scale estimate.
The AU rows are illustrative baselines, not claims about any particular
Earth--Mars conjunction ephemeris.

\begin{longtable}[]{@{}
  >{\raggedright\arraybackslash}p{(\columnwidth - 6\tabcolsep) * \real{0.2143}}
  >{\raggedleft\arraybackslash}p{(\columnwidth - 6\tabcolsep) * \real{0.2857}}
  >{\raggedleft\arraybackslash}p{(\columnwidth - 6\tabcolsep) * \real{0.2857}}
  >{\raggedright\arraybackslash}p{(\columnwidth - 6\tabcolsep) * \real{0.2143}}@{}}
\toprule\noalign{}
\begin{minipage}[b]{\linewidth}\raggedright
Illustrative separation
\end{minipage} & \begin{minipage}[b]{\linewidth}\raggedleft
Peak neutrino-carried on-slot power at 1 slot/s
\end{minipage} & \begin{minipage}[b]{\linewidth}\raggedleft
One-way vacuum flight time
\end{minipage} & \begin{minipage}[b]{\linewidth}\raggedright
Interpretation
\end{minipage} \\
\midrule\noalign{}
\endhead
\bottomrule\noalign{}
\endlastfoot
5,000 km & 24.4 MW & 0.0167 s & Terrestrial reference, same
assumptions \\
384,400 km & 144,000 MW (144 GW) & 1.28 s & Mean lunar-distance scale \\
1 AU, 149.6 million km & 21.8 billion MW (21.8 PW) & 8.32 min &
Interplanetary sensitivity example \\
2 AU & 87.4 billion MW (87.4 PW) & 16.6 min & Longer illustrative
baseline \\
\end{longtable}

At fixed slot duration, peak power scales linearly with raw slot rate.
For example, retaining 100 MW peak on-slot power and a 10 kt target at
lunar distance gives about \(100/144{,}222 \approx 0.00069\) raw
slots/s. \textbf{Each slot would last about 24 minutes; the detector
integrates throughout that on interval.} This is not a one-second flash
repeated every 24 minutes. At 1 AU, the corresponding slot duration is
about 6.9 years at 100 MW peak power. These are inversions of the
stipulated event-count model, \textbf{not useful operational rates};
coding, background accumulated over such long slots and acquisition
further reduce delivered payload. A notional 0.1 mrad divergence would
lower each tabulated power by a factor of 100, but this paper has not
demonstrated such a source at 3 GeV.

The geometry also changes the physics that the terrestrial toy model
omits. For a lunar far-side link, calculate the actual chord through the
Moon, its density profile, neutrino transmission and flavor evolution,
then test whether a receiver of the assumed mass could be deployed
there. At solar conjunction, compute the time-dependent trajectory and
impact parameter through the Sun, energy-dependent interactions,
oscillations in solar matter and background at the receiver. A path
skimming the corona and one crossing the solar interior cannot share a
universal transmission factor. Exact pointing, moving endpoints, source
duty cycle and the receiver's projected area matter at both scales.
Neutrinos travel at essentially the same vacuum speed as an unobstructed
electromagnetic signal over these baselines; occultation avoidance does
\textbf{not} remove lunar or interplanetary light time.

For an engineering comparison, define a message size, deadline and
acceptable failure probability. Compare a source-specific neutrino
link's delivered bits/s, electrical joules per delivered bit and total
receiver mass against a lunar relay or a solar-conjunction relay plus
delay-tolerant scheduling {[}23,25{]}. The NASA relay study examines
architectures to preserve Earth--Mars communications during solar
conjunction {[}25{]}. Under the present fixed-divergence example,
distance-squared dilution alone makes an interplanetary direct neutrino
link implausible at useful payload rates; any claim to the contrary
needs a demonstrated narrow source, plausible receiver and full
propagation and decoding calculation. The possible niche is a very
short, high-value signal in a constrained occultation interval, and even
that remains unproven.

\hypertarget{references}{%
\section{References}\label{references}}

{[}1{]} D. D. Stancil \emph{et al.}, ``Demonstration of Communication
using Neutrinos,'' \emph{Modern Physics Letters A} \textbf{27}, 1250077
(2012). \url{https://doi.org/10.1142/S0217732312500770} ;
\url{https://arxiv.org/abs/1203.2847}

{[}2{]} P. Huber, ``Submarine neutrino communication,'' \emph{Physics
Letters B} \textbf{692}, 268--271 (2010).
\url{https://doi.org/10.1016/j.physletb.2010.08.003} ;
\url{https://arxiv.org/abs/0909.4554}

{[}3{]} J. G. Learned, S. Pakvasa and A. Zee, ``Galactic Neutrino
Communication'' (2008). \url{https://arxiv.org/abs/0805.2429}

{[}4{]} J. Fidalgo Prieto \emph{et al.}, ``Submarine Navigation using
Neutrinos'' (2022). \url{https://arxiv.org/abs/2207.09231}

{[}5{]} S.-P. Hallsjö, ``Locating nuclear-powered submarines with
antineutrinos'' (2026). \url{https://arxiv.org/abs/2605.15642}

{[}6{]} S.-P. Hallsjö, \emph{Charged current quasi-elastic muon neutrino
interactions in the Baby MIND detector}, PhD thesis, University of
Glasgow (2018). \url{https://theses.gla.ac.uk/41123/}

{[}7{]} Particle Data Group, ``Neutrino Cross Section Measurements,''
\emph{Review of Particle Physics} (2025).
\url{https://pdg.lbl.gov/2025/reviews/rpp2025-rev-nu-cross-sections.pdf}

{[}8{]} nuSTORM Collaboration, ``Neutrinos from Stored Muons (nuSTORM)''
(2025). \url{https://arxiv.org/abs/2505.06137}

{[}9{]} P. B. Denton and R. Pestes, ``Neutrino Oscillations through the
Earth's Core,'' \emph{Physical Review D} \textbf{104}, 113007 (2021).
\url{https://doi.org/10.1103/PhysRevD.104.113007} ;
\url{https://arxiv.org/abs/2110.01148}

{[}10{]} DUNE Collaboration, \emph{Deep Underground Neutrino Experiment,
Far Detector Technical Design Report, Volume I: Introduction to DUNE}
(2020). \url{https://arxiv.org/abs/2002.02967}

{[}11{]} S.-P. Hallsjö, \emph{Direct through-Earth neutrino
communication: reproducible scoping study}, analysis scripts, data and
figures archived in the repository snapshot at commit
7dc09e25e9ad65383ff19733ab7f88fd38aa7b9d (2026). Reproduction checkout
and commands are listed in the repository README.

{[}12{]} Hyper-Kamiokande Proto-Collaboration, \emph{Hyper-Kamiokande
Design Report} (2018). \url{https://arxiv.org/abs/1805.04163}

{[}13{]} J. I. Statman, ``Optimizing the Galileo Space Communication
Link,'' \emph{Interplanetary Network Progress Report} \textbf{42-116},
114--120 (1994).
\url{https://ipnpr.jpl.nasa.gov/progress_report/42-116/116k.html}

{[}14{]} B. Moision and J. Hamkins, ``Deep-Space Optical Communications
Downlink Budget: Modulation and Coding,'' \emph{Interplanetary Network
Progress Report} \textbf{42-154}, 1--28 (2003).
\url{https://ipnpr.jpl.nasa.gov/progress_report/42-154/154K.html}

{[}15{]} J. I. Statman, K.-M. Cheung, T. H. Chauvin, J. Rabkin and M. L.
Belongie, ``Decoder synchronization for deep space missions'' (1994),
NASA Technical Reports Server ID 19940025161.
\url{https://ntrs.nasa.gov/citations/19940025161}

{[}16{]} A. W. Sáenz, H. Uberall, F. J. Kelly, D. W. Padgett and N.
Seeman, ``Telecommunication with neutrino beams,'' \emph{Science}
\textbf{198}, 295--297 (1977).
\url{https://doi.org/10.1126/science.198.4314.295}

{[}17{]} M. Karzand and L. R. Varshney, ``Communication Strategies for
Low-Latency Trading'' (2015). \url{https://arxiv.org/abs/1504.07227}

{[}18{]} G. Laughlin, A. Aguirre and J. Grundfest, ``Information
Transmission Between Financial Markets in Chicago and New York'' (2013).
\url{https://arxiv.org/abs/1302.5966}

{[}19{]} Corning, \emph{SMF-28e+ Photonic Optical Fiber} product
information, effective group index 1.4677 at 1550 nm.
\url{https://www.corning.com/media/worldwide/csm/documents/Corning\%20SMF28e\%2B\%C2\%AE\%20Photonic\%20Specialty\%20Fiber.pdf}

{[}20{]} Consultative Committee for Space Data Systems, \emph{Lossless
Data Compression}, CCSDS 121.0-B-3 (2020).
\url{https://ccsds.org/Pubs/121x0b3.pdf}

{[}21{]} Consultative Committee for Space Data Systems, \emph{TM
Synchronization and Channel Coding}, CCSDS 131.0-B-6 (2026).
\url{https://ccsds.org/view/bluebooks/entry/4803/}

{[}22{]} Y. Collet and M. Kucherawy, ``Zstandard Compression and the
`application/zstd' Media Type,'' RFC 8878 (2021).
\url{https://datatracker.ietf.org/doc/html/rfc8878}

{[}23{]} NASA, ``Exploration and Space Communications: Lunar
Communications Relay and Navigation Systems.''
\url{https://www.nasa.gov/goddard/esc/lcrns/}

{[}24{]} NASA/JPL, ``What's Mars Solar Conjunction, and Why Does It
Matter?'' (2019).
\url{https://www.jpl.nasa.gov/news/whats-mars-solar-conjunction-and-why-does-it-matter/}

{[}25{]} NASA Technical Reports Server, ``Lagrange-Based Options for
Relay Satellites to Eliminate Earth-Mars Communications Outages During
Solar Superior Conjunctions'' (2020).
\url{https://ntrs.nasa.gov/citations/20205007788}

\end{document}